\documentclass{article}
\usepackage{graphicx} 
\usepackage{tabularx}
\usepackage{multirow}
\usepackage{booktabs} 
\usepackage[table,xcdraw]{xcolor}  
\usepackage[a4paper, margin=1in]{geometry} 
\usepackage{amsmath} 
\usepackage[colorlinks=true,
            linkcolor=blue,
            citecolor=blue,
            urlcolor=blue]{hyperref}
\usepackage{float}
\usepackage[utf8]{inputenc}
\usepackage{float}
\usepackage{subcaption} 
\usepackage{amsmath, amssymb}
\usepackage{listings}  
\usepackage{xcolor}    
\usepackage[style=apa]{biblatex} 
\DeclareCaptionLabelSeparator{boldcolon}{\textbf{:}\enspace}
\DeclareCaptionFormat{custom}{\textbf{#1#2}#3}
\title{Animated Territorial Data Extractor (ATDE): A Computer-Vision Method for Extracting Territorial Data from Animated Historical Maps
}

\author{
    \textbf{Hamza Alshamy}, \quad 
    \textbf{Isaiah Woram}\textsuperscript{*}, \quad
    \textbf{Advay Mishra}\textsuperscript{*},  \\
    \quad \textbf{Zihan Xia}\textsuperscript{*}, \quad
    \textbf{Pascal Wallisch} \\
    Center for Data Science, New York University \\
    \texttt{\{ha2486, idw2005, am11369, zx1117, pw44\}@nyu.edu}
}

\date{} 

\begin{document}

\maketitle

\begingroup
\renewcommand\thefootnote{*}
\footnotetext{Work completed while at the Center for Data Science, New York University.}
\endgroup

\begin{abstract}
We present \textit{Animated Territorial Data Extractor (ATDE)}, a computer vision tool that extracts quantitative territorial data from animated historical map videos. ATDE employs HSV-based color segmentation, RGB channel filtering, and Direct-Neighbor Filtering to identify and count pixels representing territorial control. Combined with preprocessing for temporal alignment and cross-video scaling, the pipeline converts animated videos into structured time-series data. We demonstrate the tool on ten Chinese dynasties (200 BCE – 1912 CE), producing year-by-year pixel counts that align with expected historical patterns. While not a substitute for authoritative historical datasets, ATDE is well-suited for educational demonstrations, preliminary data exploration, and comparative analysis of territorial dynamics. The tool requires no pre-existing shapefiles and can be applied to any animated map video given seed colors and basic configuration. Code and examples are available on \href{https://github.com/hamzaalshamy/animated-territorial-data-extractor.git}{GitHub}.
\end{abstract}

\section{Introduction}

Animated “history-of-the-world” videos have become a popular medium for disseminating historical narratives, captivating millions of viewers and enriching classroom discussions of empire formation, territorial conflicts, and state collapse. Yet, despite their widespread availability and pedagogical value, these animated visualizations remain largely untapped as resources for rigorous scholarly inquiry. A key obstacle is that the territorial information underlying these videos is rarely available in structured, analytically usable form. Typically, creators paint these maps frame‐by‐frame using consumer software such as MS Paint, exporting the final product as a rasterized MP4 file without accompanying data on borders or geographic attributes. Even when animators do utilize professional Geographic Information System (GIS) tools to produce these visuals, the resulting structured files are not always shared publicly. Consequently, a vast repository of detailed spatio-temporal data exists openly online yet remains effectively inaccessible for quantitative historical, economic, and political research.

Animated map videos represent an accessible but underutilized source of territorial information. While authoritative historical datasets exist in academic repositories and GIS databases, these videos offer a complementary resource, particularly for analyzing how territorial changes are visualized and communicated in popular media. Currently, the quantitative information embedded in these videos remains locked in rasterized bitmap format, inaccessible for systematic analysis. Video creators lack tools to validate the accuracy of their animations, educators cannot easily extract data for teaching quantitative methods, and researchers interested in studying how historical narratives are visually represented have no automated way to analyze these increasingly popular media artifacts. If the territorial data depicted in animated map videos could be reliably extracted, it would enable new forms of analysis: validating video accuracy against established sources, comparing how different creators represent the same historical periods, and providing rapid preliminary datasets for educational purposes or prototyping digitization projects. In this paper, we introduce the Animated Territorial Data Extractor (ATDE), an automated computer vision pipeline that extracts per-frame counts of pixels depicting territorial control from rasterized animations. ATDE converts previously inaccessible video content into structured data compliant with FAIR (Findability, Accessibility, Interoperability, and Reusability) principles (\cite{Wilkinson2016FAIR}).

To demonstrate the utility of the extractor and its potential application of the data, we apply the pipeline to extract annual territorial extents for ten Chinese dynasties (200 BCE – 1912 CE), all depicted on a consistent, shared basemap. The resulting time series closely align with established historiographic periodizations. This illustrative case study is intentionally modest in scope; its primary purpose is not to offer a comprehensive reinterpretation of imperial Chinese history but rather to demonstrate the tool's functionality and show how extracted data can be used for comparative analysis of territorial dynamics. Potential applications include examining patterns of territorial stability and collapse, comparing expansion rates across different historical periods, and providing preliminary datasets for educational demonstrations or exploratory analysis—uses that benefit from systematic extraction of video content even when authoritative historical datasets exist.

\section{Methodology}

Animated videos showing year-by-year territorial expansion of empires and dynasties typically do so by representing territory with a unique color. As territory expands in a given year, the number of pixels representing that territory increases, and as territory shrinks, so too does the number of pixels representing that territory. We take advantage of this mapping by extracting the number of pixels representing territory (in a certain RGB color) and outputting a dictionary containing the number of those pixels for each year. It does so by taking a list of RGB colors as input, and, for each frame, counts pixels matching the specified colors, aggregating them by year to return a \{year: pixel\_count\} dictionary.

As an overview, the methodology consists of two main parts. First, video pre-processing, including cropping the duration and spatial frame of a video, as well as removing redundant or invalid frames from the video. This process produces a shorter, condensed video in which each spatiotemporally cropped frame presents new and valid information about territorial control. The second part is the ATDE, which takes this condensed video and detects the number of pixels matching the color criterion representing territory. Furthermore, the function takes on hyperparameters controlling the effects of embedded robustness checks (e.g., color channel restrictions and noise cancellation).

\subsection{Video Pre-processing}

In order to use the information contained in the raw videos, we have to extract the individual frames. This will happen in the next step detailed below. A necessary preliminary to this is to first convert the raw video to usable - cropped - regions, both spatially and temporally. The reason for this necessity lies in the fact that our algorithms will interpret all pixels and their color values in a meaningful way. Thus, we have to constrain them to regions where that is the case, in other words where the data is valid. Specifically, in the temporal domain, many videos contain an intro/outro - usually some channel- and/or video-specific material - that needs to be chopped off. This can be done by hand/visual inspection or algorithmically by finding the frame of large transitions in pixel values (frame by frame). Once the video transitions to the "meat" of the video (the evolution of the map), most pixel values are constant, frame-by-frame, with only relatively minor changes. Similarly, the raw frames typically a large amount of - for the purposes of this enterprise - extraneous information, such as the name of the dynasty, a picture of the ruler, etc. For the purpose of our analysis, we will focus on and utilize information from 3 parts of the raw frame: The map, the clock and the color scheme, see Figure~\ref{fig:schematic} for a schematic of this. 

\paragraph{Temporal Alignment ("Frame Clock").} In order to use the information contained in the video, it is critical to ensure that a given time step in the video - be it a frame or multiple frames - corresponds to a constant amount of years passing in history and consequently taking up the same space on x-axis. In other words, we have to ensure the consistency of the clock throughout the video. We cannot simply assume that the clock has the same speed throughout the video, even if it might look innocuous and constant at first glance. Instead, we have to validate the integrity of the “frame clock” throughout the video. We do so here by “delta encoding” to construct a consistent “frame clock”. Specifically, we use the information encoded in the "Clock" in Figure~\ref{fig:schematic} and count the number of black pixels (that represent the year) in front of the lighter (gray) background within this section. The passage of time in a video will manifest as a change in the number of these pixels, as, for example, the number “9” has a different number of black pixels as the number “0”.

Our goal is a one-frame-per-year series so that the $x$-axis in this graph reflects years, independent of editorial pacing. We build the frame-to-year mapping by detecting digit changes inside the clock window $\mathcal{C}$.

\begin{figure}[H]
    \centering
    \includegraphics[width=.8\textwidth]{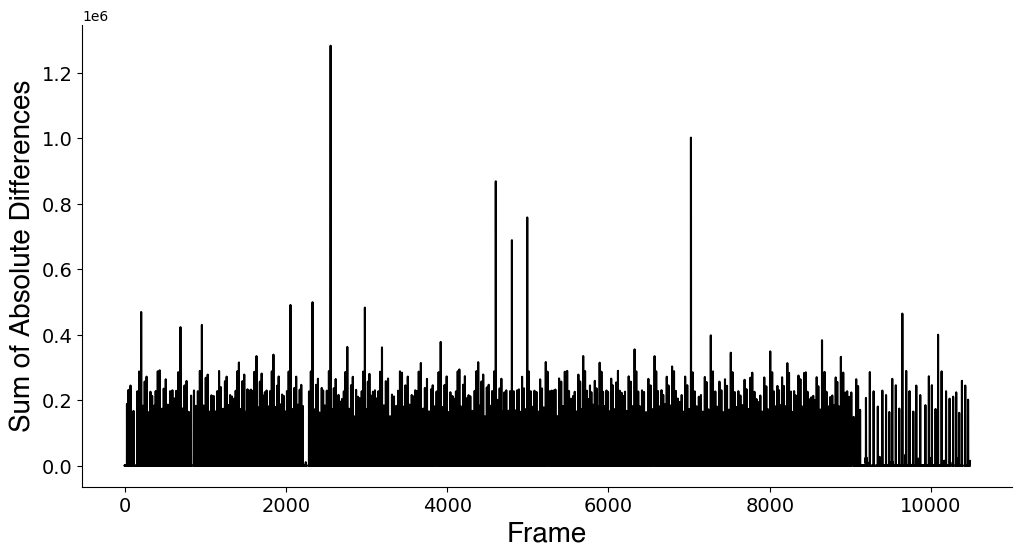}
    \caption{Frame clock transitions. The x-axis indicates the frames throughout the preprocessed video. The y-axis depicts the sum of the absolute differences in RGB values across adjacent frames. The black trace represents this sum of absolute differences throughout the preprocessed video.}
    \label{fig:sum_of_abs}
\end{figure}

Let $I_t$ denote frame $t$. We compute a per-frame difference score over $\mathcal{C}$,
\[
S_t \;=\; \sum_{(x,y)\in \mathcal{C}} \big\lVert I_t(x,y) - I_{t-1}(x,y) \big\rVert_1,
\]
which spikes when the on-screen digits change. We choose a threshold $\tau$ from the histogram of $\{S_t\}$ (visually or via a rule) and treat indices with $S_t>\tau$ as change points. For each interval between change points we retain exactly one representative frame and assign it to the corresponding year by reading the digits in $\mathcal{C}$. This yields a list of frame indices $\{f_y\}_{y=\text{start}}^{\text{end}}$ containing exactly one frame per year and removes redundant frames automatically (see Figure~\ref{fig:sum_of_abs} for an example of this method).

The spikes in this graph represent changes in the year. In addition, due to visual noise, there will always be a subtle change of pixel values in this window (1d), even if the year has not changed. Thus, it is important to set a threshold that takes this noise level into account - the sum of the absolute differences of pixel values in this window. Here, we set this threshold by visual inspection to 50,000 (the sum of the absolute values of pixel changes and validated this manually). We recommend determining this threshold by looking at the histogram of the sum of these absolute differences. Implementing this method allows us to extract only one frame per year across the entire range of the video, regardless of whether the clock is consistent or not. This allows us to interpret the x-axis of the final graph (see Figures~\ref{fig:Song Cycle} and~\ref{fig:ATDE_full}) in terms of years passed. An added benefit of this approach is that all redundant frames - often a given year is represented by a varying number of multiple frames - are removed en passant, yielding a much condensed video that is much more efficient to process in subsequent steps. 

\subsection{The Extractor Function}

The purpose of ATDE function is to read a video, and for each frame, identify areas that match specified colors representing territory controlled by a particular dynasty and count the pixels of that color, and eventually, plot the number of pixels across time. This function takes parameters, including: \texttt{video\_path}, \texttt{output\_path} (for validation video\footnote{The validation video is described in Section~\ref{frame handling}}), \texttt{main\_colors} (list of RGB shades representing territory), \texttt{start\_year} (of the dynasty), and \texttt{end\_year}.

\begin{figure}[H]
    \centering
    \includegraphics[width=.8\textwidth]{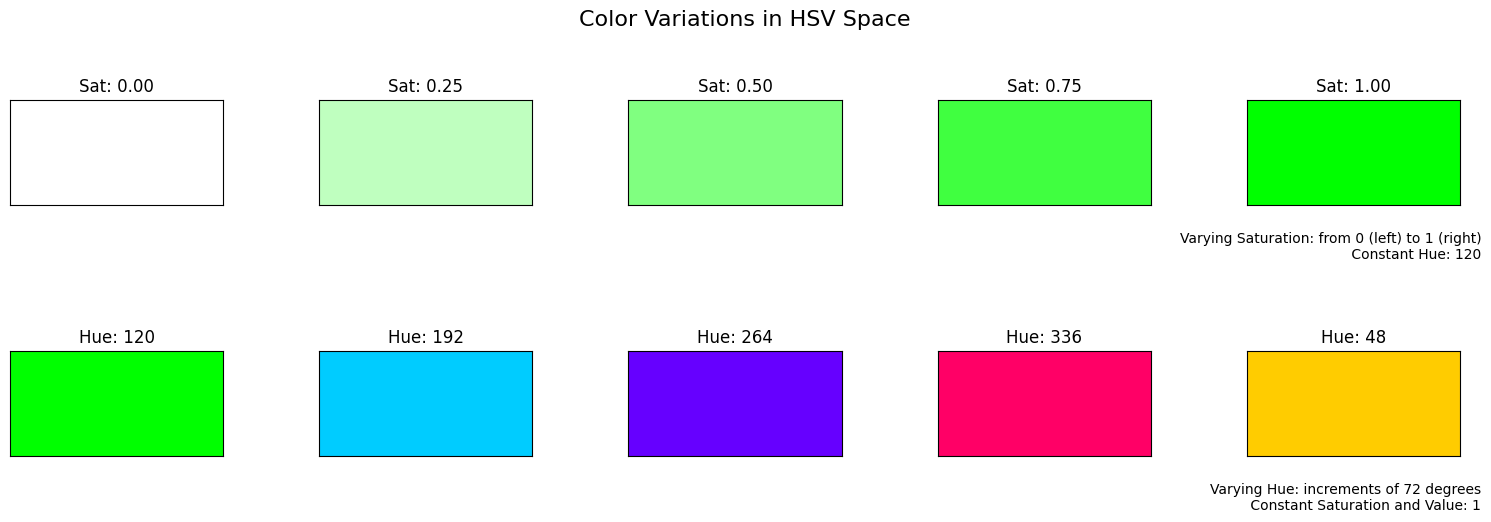}
    \caption{Demonstration of HSV color space properties. Top: varying saturation produces different shades of the same hue. Bottom: varying hue produces distinct colors. This separation enables ATDE to capture multiple shades of dynasty colors while maintaining specificity.}
    \label{fig:hsvColor}
\end{figure}

Whereas the input space of the videos represents colors as a tuple of RGB (red, green, blue) values, we perform a RGB to HSV transformation to better combine and separate colors that correspond to particular dynasties. HSV (hue, saturation, value) space allows for better color segmentation than RGB, as each dynasty is represented by several shades of the same hue in the videos. As shown in Figure~\ref{fig:hsvColor} (top row), multiple shades of the same hue can be represented in HSV space by varying saturation, enabling efficient color segmentation for territorial analysis.

The function utilizes the \texttt{cv2} module from \texttt{OpenCV} library and starts by reading the condensed input video produced in the first part, and loads the video as a video capture object.

\subsubsection{Color Detection Set-up}
There are several ways to detecting the color (and corresponding pixels) representing the territory controlled by a given dynasty. For example, an overly simplistic approach would be to only count the pixels with exactly matching RGB  values. However, due to scatter in the exact color RGB values of the pixels, such an approach may not capture all the desired pixels, undercounting the true value. Instead, this function applies an HSV color range within which it detects desired pixels representing territory. Thus, the criterion for color detection is not that each pixel meets an \textit{exact} color value, but that each pixel falls within a certain \textit{range} of values.

At the same time, while this approach allows to handle and count different shades of a given hue as the same and mapping it to a particular color, as well as inherently handle color noise, this range needs to be carefully calibrated, so as not to be overly inclusive to avoid overcounting. 

To do so, first the list of RGB input values \footnote{The parameter, \texttt{main\_colors} accepts a list which allows the detection of multiple shades of the same color instead of the exact RGB value.} is converted to BGR (blue, green, red) - the default color space in \texttt{cv2} - and then to HSV.

Second, the lower and upper HSV ranges are defined by three hyperparameters: \texttt{hsv\_range}, \texttt{lower\_sv}, and \texttt{upper\_sv}. \texttt{hsv\_range}  describes how much variation in the hue is allowed for each HSV color to be considered as the same (default = $\pm 10$). \texttt{lower\_sv} and \texttt{upper\_sv} indicates the variation in saturation and value respectively; the lower the saturation and value in \texttt{lower\_sv} (default = 100 \footnote{The range of saturation and value in HSV are typically between 0-100\%. However, the range for them in \texttt{cv2} is between 0-255}), the more pale shades of the hue will be captured and counted as the same, and \texttt{upper\_sv} is set to the of max 255 by default to capture the most vivid colors of the hue. 

Thus, for each HSV color, a lower and upper bound is defined by these three hyperparameters. For example, if the list of RGB input values contains three colors, this process will yield three HSV lower colors and three upper colors that are mapped onto the same color.

\subsubsection{Frame Handling and Pixel-count Extraction}
\label{frame handling}
\begin{figure}[htbp]
    \centering
    \begin{subfigure}[b]{0.45\textwidth}
        \centering
        \includegraphics[width=\textwidth]{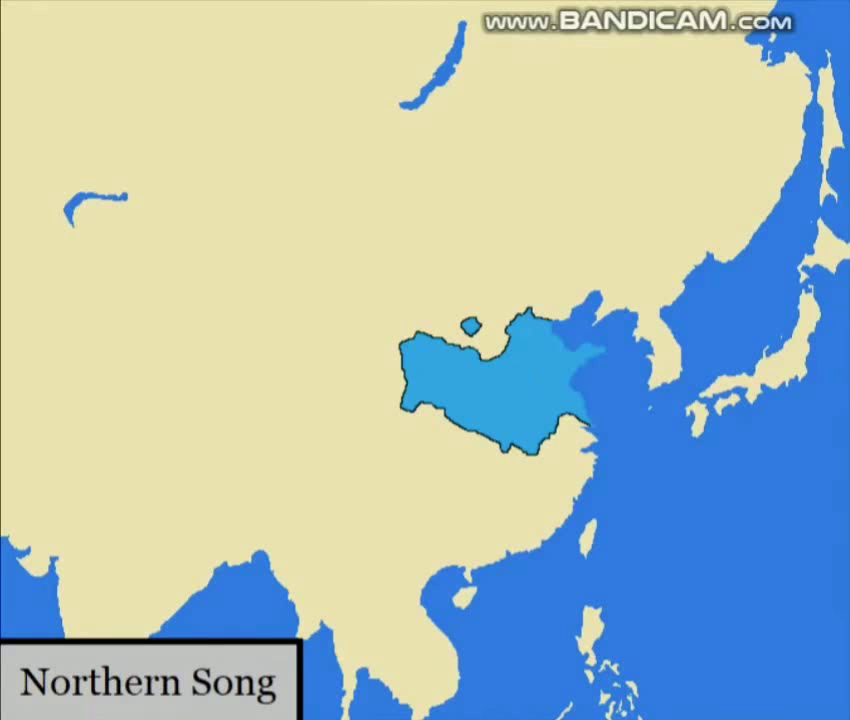}
        \caption{Frame of Original Video}
        \label{fig:song_1}
    \end{subfigure}
    \hfill
    \begin{subfigure}[b]{0.45\textwidth}
        \centering
        \includegraphics[width=\textwidth]{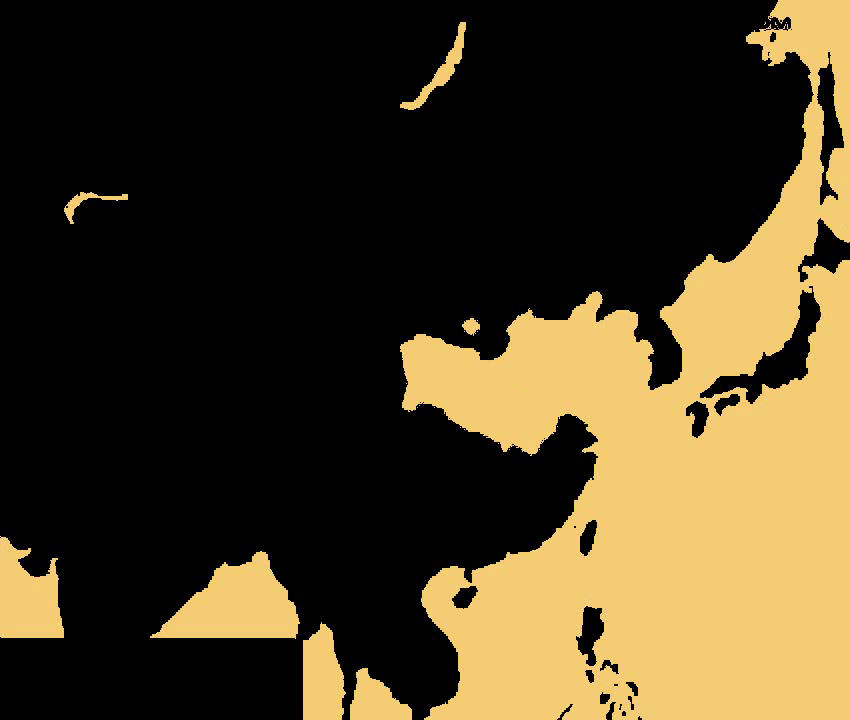}
        \caption{Frame of Validation Video}
        \label{fig:song_2}
    \end{subfigure}

    \vspace{1em} 

    \begin{subfigure}[b]{0.45\textwidth}
        \centering
        \includegraphics[width=\textwidth]{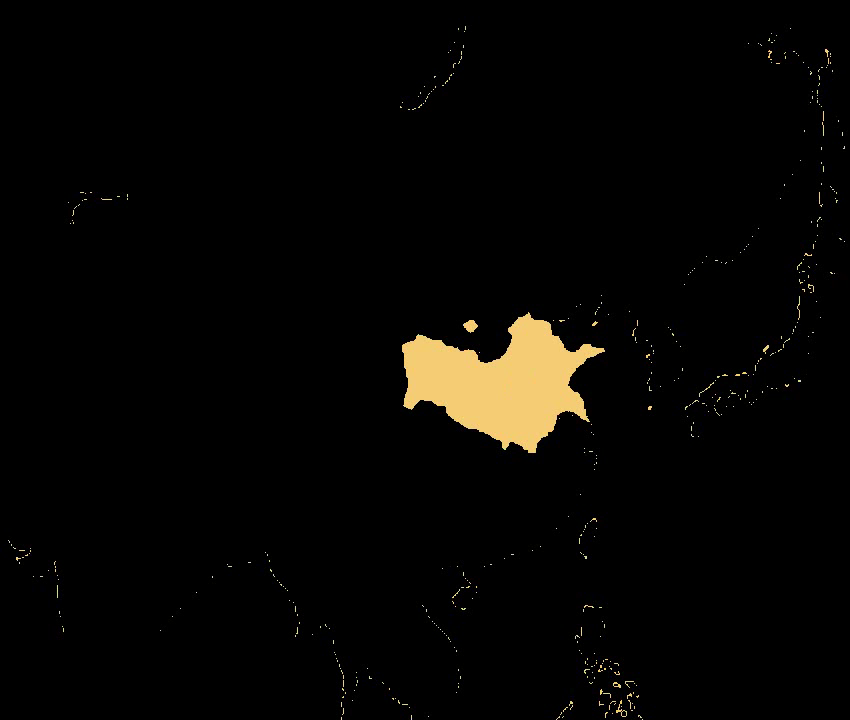}
        \caption{Frame of Validation Video with RGB Restriction (\( \text{green} < 150 \))}
        \label{fig:song_3}
    \end{subfigure}
    \hfill
    \begin{subfigure}[b]{0.45\textwidth}
        \centering
        \includegraphics[width=\textwidth]{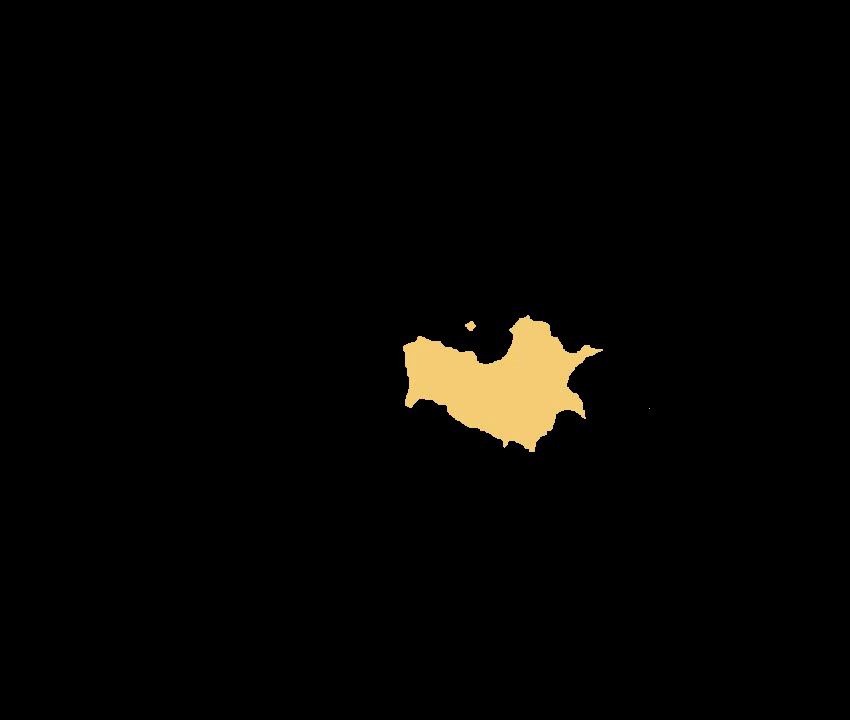}
        \caption{Frame of Validation Video with RGB restriction and DNF Function Applied (8 neighbors)}
        \label{fig:song_4}
    \end{subfigure}

    \caption{Color Segmentation Checks. (a) A frame from the original video, where the goal is to extract the number of pixels representing the dynasty’s territory (light blue). (b) The corresponding validation frame produced by color detection; however, because the ocean shares a similar hue, pixels outside the territory (ocean) are incorrectly counted. (c) To address this, we introduce RGB channel restrictions. Here, we exclude pixels with a green value below 150, which effectively removes most of the ocean, though some border noise remains. (d) Applying the Direct Neighbor Filtering (DNF) function with 8-neighbor connectivity eliminates this residual noise, leaving only the targeted territory.}
    \label{fig:panel}
\end{figure}

Extracting the quantitative information from each frame of the video is a four-step process. Briefly, each frame is read in, converted to HSV, and then all pixels of a frame are classified as part of the territory controlled by the dynasty or not, as laid out above. To ensure specificity, an optional RGB channel color restriction as well as \textit{Direct Neighbor Filtering (DNF)} is also applied to remove edge artifacts. See Figure~\ref{fig:panel} for the effect of each step on color segmentation and detection.

First, for each frame that is read from the video, its pixel values are converted from RGB to HSV. Afterwards, a binary mask for the current frame is initialized (i.e. all values in the mask are zero).

Second, each pair of lower/upper HSV bounds are iterated by using a \texttt{zip()} function. Using \texttt{cv2.inRange}, the pixels that fall within the range of lower/upper bounds are detected in a new color mask; if a pixel in the frame is within the range, it is set to white (True\footnote{Pixel value: 255}) while others are set to black (False\footnote{Pixel value: 0}). Using \texttt{cv2.bitwise\_or} with the initialized mask and color mask, if a pixel in either of the masks is set to True, then the pixel is updated to True in the previously initialized mask. In the case of three pairs of lower/upper HSV bounds, we see if a pixel in a frame falls within the bounds of a pair, once this process is done, all pixels in the frame that fell within the bounds are set to True in a colored mask. All that did not are set to False. The same process is applied to the same frame but with the second HSV bounds. If a pixel is True in either of the now-written-over initialized mask or colored mask, then the now-written-over initialized mask is updated. The process is applied to all HSV bounds for all frames.

For each frame, a \emph{validation frame} is produced and accumulated to produce a \emph{validation video}. A validation video is a visual representation that provides a frame-by-frame visualization of which areas in the original video were detected as matching the specified color criteria. It is a visual representation of the pixels that are counted through the analysis which allows for visual assessment of the accuracy of the color detection algorithm. One can see if the algorithm is correctly identifying the areas of interest and if there are any false positives or negatives. Each frame in the validation video essentially contains two types of information: highlighted areas that match the color detection criteria and black areas that do not match the criteria. For example, as can be seen in Figure~\ref{fig:song_2}, the color detection algorithm detected the highlighted territory area represented in Figure~\ref{fig:song_1}, however, it also detected bodies of water due to the similar pixel values (false positives). Thus, the validation video offers feedback on how the algorithm is working.

Third, in the case that the function is detecting the background color, ocean color, or any non-territory color due to color similarity, an RGB channel boundary can be applied. This color filtering technique allows the separation of similar colors. For example, observe the color similarity in the Song Dynasty territory and ocean colors in Figure~\ref{fig:song_1}. The validation frame in Figure~\ref{fig:song_2} shows that the function is detecting the territory as well as the ocean color. If there is an RGB color channel in which the territory and ocean (or non-territory) colors do not overlap, a boundary can be set. In the case of the Song Dynasty, there is no overlap in the green channel between the territory colors and the ocean color. Specifically, the green channel values for the Song Dynasty territory range from 170 to 238, while the ocean's green channel value is 135 (refer to Table~\ref{tab:rgb_values}). Therefore, from the detected pixels, a green channel filtering restriction can be applied to only keep pixels whose green channel value is less than 150\footnote{The reason for not setting the green channel restriction to exactly 136 is to account for variation in color.}. Figure~\ref{fig:song_3} illustrates the significant effect of the filtering method.

\begin{table}[h]
\centering

\begin{minipage}{.5\textwidth}
\centering
\begin{tabular}{|c|c|}
\hline
\textbf{Song Dynasty RGB Values} & \textbf{Color} \\ \hline
(47, 170, 235) & \cellcolor[rgb]{0.18,0.67,0.92} \\ \hline
(103, 207, 254) & \cellcolor[rgb]{0.40,0.81,1.00} \\ \hline
(201, 238, 254) & \cellcolor[rgb]{0.79,0.93,1.00} \\ \hline
\end{tabular}
\subcaption{Colors of the Song Dynasty}
\label{tab:song_colors}
\end{minipage}%
\begin{minipage}{.5\textwidth}
\centering
\begin{tabular}{|c|c|}
\hline
\textbf{Ocean RGB Value} & \textbf{Color} \\ \hline
\multirow{3}{*}{(49, 135, 235)} & \cellcolor[rgb]{0.19,0.53,0.92} \\ \cline{2-2}
 & \cellcolor[rgb]{0.19,0.53,0.92} \\ \cline{2-2}
 & \cellcolor[rgb]{0.19,0.53,0.92} \\ \hline
\end{tabular}
\subcaption{Ocean Color}
\label{tab:ocean_color}
\end{minipage}

\caption{Comparison of RGB Values for Song Dynasty and Ocean Colors. (a) The Song Dynasty territory is represented by three shades of blue with green channel values ranging from 170 to 238. (b) The ocean color has a similar hue but a distinct green channel value of 135. This non-overlapping green channel distribution enables RGB channel filtering to separate territory pixels from ocean pixels during extraction (see Figure~\ref{fig:song_3} for the effect of applying a green channel threshold of 150).}
\label{tab:rgb_values}

\end{table}

Lastly, the final filtering technique in the frame processing step is Direct Neighbor Filtering (DNF) whose goal is to remove isolated pixels and reduce noise. Simply, when DNF is applied to a frame (matrix), the immediate surrounding pixels of a point that are detected are counted, and points whose detected neighbors are fewer than a minimum threshold are removed. Thus, removing isolated pixels. The matrices in Figure~\ref{fig:dnf_process} show the effect of DNF on a frame (matrix).

\begin{figure}[ht!]
\centering
\[
\begin{array}{cccccc}
\begin{bmatrix}
0 & 1 & 0 & 0 & 1 \\
1 & 1 & 0 & 1 & 0 \\
0 & 1 & 1 & 1 & 0 \\
0 & 0 & 1 & 0 & 0 \\
1 & 0 & 0 & 0 & 1 \\
\end{bmatrix}
& \xrightarrow{\text{Count Neighbors}}
& \begin{bmatrix}
3 & 2 & 3 & 2 & 1 \\
3 & 4 & 6 & 3 & 3 \\
3 & 4 & 5 & 3 & 2 \\
2 & 4 & 3 & 4 & 2 \\
0 & 2 & 1 & 2 & 0 \\
\end{bmatrix}
& \xrightarrow{\substack{\text{Apply Threshold} \\ \texttt{(min\_neighbors = 3)}}}
& \begin{bmatrix}
0 & 0 & 0 & 0 & 0 \\
0 & 1 & 1 & 1 & 0 \\
0 & 1 & 1 & 1 & 0 \\
0 & 1 & 0 & 1 & 0 \\
0 & 0 & 0 & 0 & 0 \\
\end{bmatrix}
\end{array}
\]
\caption{The Effect of Direct Neighbor Filtering on a \(5 \times 5\) matrix}
\label{fig:dnf_process}
\end{figure}

Essentially, DNF is a convolution operation on a frame with a \(3 \times 3\) kernel. When DNF is applied to a frame, a binary mask of the frame is created in which positive pixel values are set to 1, otherwise 0. The 1/0 binary configuration of the mask is then converted to 8-bit unsigned integers (255/0). The \(3 \times 3\) kernel is of all ones and zero in the center. The design of the kernel allows for counting the number of detected pixels of the 8 neighboring pixels of a point. The hyperparameter, \texttt{min\_neighbors}, which sets the minimum threshold for the required number of detected neighboring pixels has an effective maximum value of 8. Since the kernel only considers the immediate surrounding pixels of a point, setting the minimum number of neighbors to 9, for example, will return 0 on all pixels. Once the minimum number of neighbors is defined \footnote{default \texttt{min\_neighbors = 5}}, a convolution is applied to the frame using \texttt{cv2.filter2D} in which the kernel slides over each pixel, counting the number of detected pixels of the 8 neighboring pixels. Lastly, from the originally detected pixels (those with value 255), only those at or above the minimum number of detected neighbors remain in the filtered frame, set to 255, while all others are set to 0.

Most of the analysis of the ATDE is completed in the frame processing step. The step starts by reading and converting a frame from RGB to HSV, applying the frame processing algorithm, including the optional RGB channel boundary and DNF. Afterward, create a validation frame to create a validation video. These steps are repeated as many times as there are frames in the input video. The number of pixels representing a dynasty’s territory is stored in a dictionary, where each entry corresponds to the pixel count for a given year. Since each year in our condensed videos directly maps to a single year, the dictionary effectively tracks the dynasty’s territorial extent over time. 

\subsubsection{The Journey of a Frame}

Now that the main components (color-detection setup, frame processing, and filtering techniques) of the ATDE function are described, the journey of a frame can now be told. Each frame goes through a loop in which it is processed and a validation frame is created. The accumulation of validation frames creates the validation video.

Looping through each frame of a video, a frame is read, and a mask for the current frame is created. The mask applies the steps outlined in the frame processing section. That is, the detected pixels are those that fall within the HSV upper and lower bounds. An optional RGB color channel restriction is applied to the mask, and lastly, DNF is applied to reduce noise. Once these steps are completed for a frame, the non-zero pixel count for the frame is appended to a dictionary at the appropriate year interval. Lastly, a validation frame is created. The validation frame starts as an empty frame with the same dimensions as the original video, and the pixels in the validation frame that correspond to non-zero pixels in the mask are set to the mean of the input RGB values. The validation frame is written to an output video file.

\section{Use-case: Analysis of Chinese Dynasties}

To demonstrate the functionality of ATDE, we applied this pipeline to extract territorial trajectories for ten major Chinese dynasties spanning over two millennia (200 BCE – 1912 CE). All animations were sourced from videos sharing a consistent basemap, enabling meaningful cross-dynasty comparisons after appropriate scaling adjustments.

\begin{figure}[H]
    \centering
    \includegraphics[width=\textwidth]{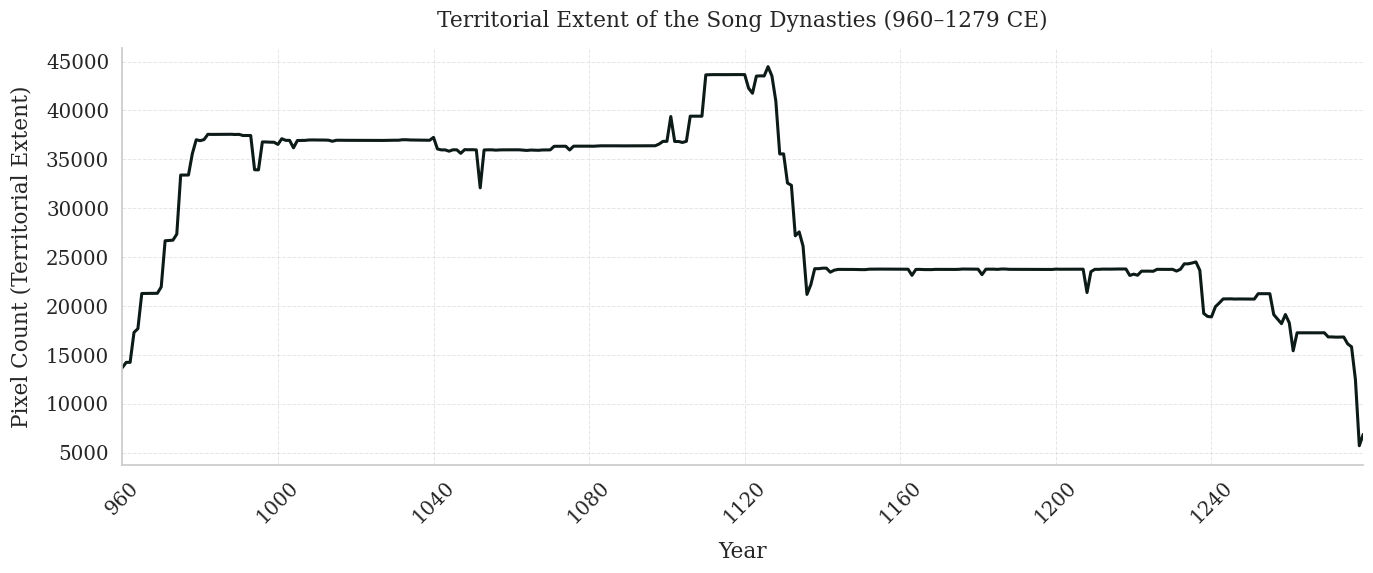}
    \caption{Territorial Extent of the Song Dynasty (960–1279 CE). Extracted pixel counts show rapid initial expansion, a major territorial loss around 1127, and gradual decline through the 13th century. The trajectory aligns with known historical periodization of the Northern Song (960–1127) and Southern Song (1127–1279) periods.}
    \label{fig:Song Cycle}
\end{figure}

\paragraph{Single Dynasty Extraction.} Figure~\ref{fig:Song Cycle} illustrates the extracted territorial trajectory for the Song Dynasty. The x-axis represents years, while the y-axis shows the raw pixel count: The number of pixels matching the Song Dynasty's color specification in each frame after applying HSV-based segmentation, RGB channel filtering, and Direct Neighbor Filtering. The resulting curve captures the dynasty's territorial evolution, including its initial consolidation, periods of relative stability, and eventual contraction. This single-dynasty analysis demonstrates ATDE's ability to convert animated visualizations into quantifiable time-series data.

\paragraph{Multi-Dynasty Comparison.} Comparing territorial extents across multiple dynasties requires addressing two key challenges: differing map scales and varying absolute territorial sizes.

\subparagraph{(i) Scale Normalization.} Video creators occasionally zoom in or out depending on the geographic scope relevant to a particular historical period. Even when videos share a common basemap, these zoom changes alter the pixel-to-area ratio, making raw pixel counts non-comparable across videos. To address this, we leverage invariant geographic features—specifically, bodies of water—to compute scale factors that normalize pixel counts across different videos. This cross-video scaling procedure is detailed in Appendix~\ref{Cross-video Scaling}.

\subparagraph{(ii) Relative Normalization.} Even after spatial scaling, different dynasties controlled vastly different maximum territorial extents: Comparing the raw scaled pixel count of a smaller regional dynasty to a continent-spanning empire would obscure meaningful patterns. To enable visual comparison of territorial dynamics regardless of absolute size, we normalize each dynasty's scaled trajectory by dividing by the global maximum pixel count across all dynasties and all time periods. This transforms all trajectories to a common scale where the largest territorial extent ever achieved (across any dynasty) equals 1. This reinterpretation of the y-axis allows us to compare the relative patterns of expansion, consolidation, and contraction across dynasties with different geographic footprints, while preserving the proportional relationships between their absolute sizes.

\begin{figure}[H]
    \centering
    \includegraphics[width=\textwidth]{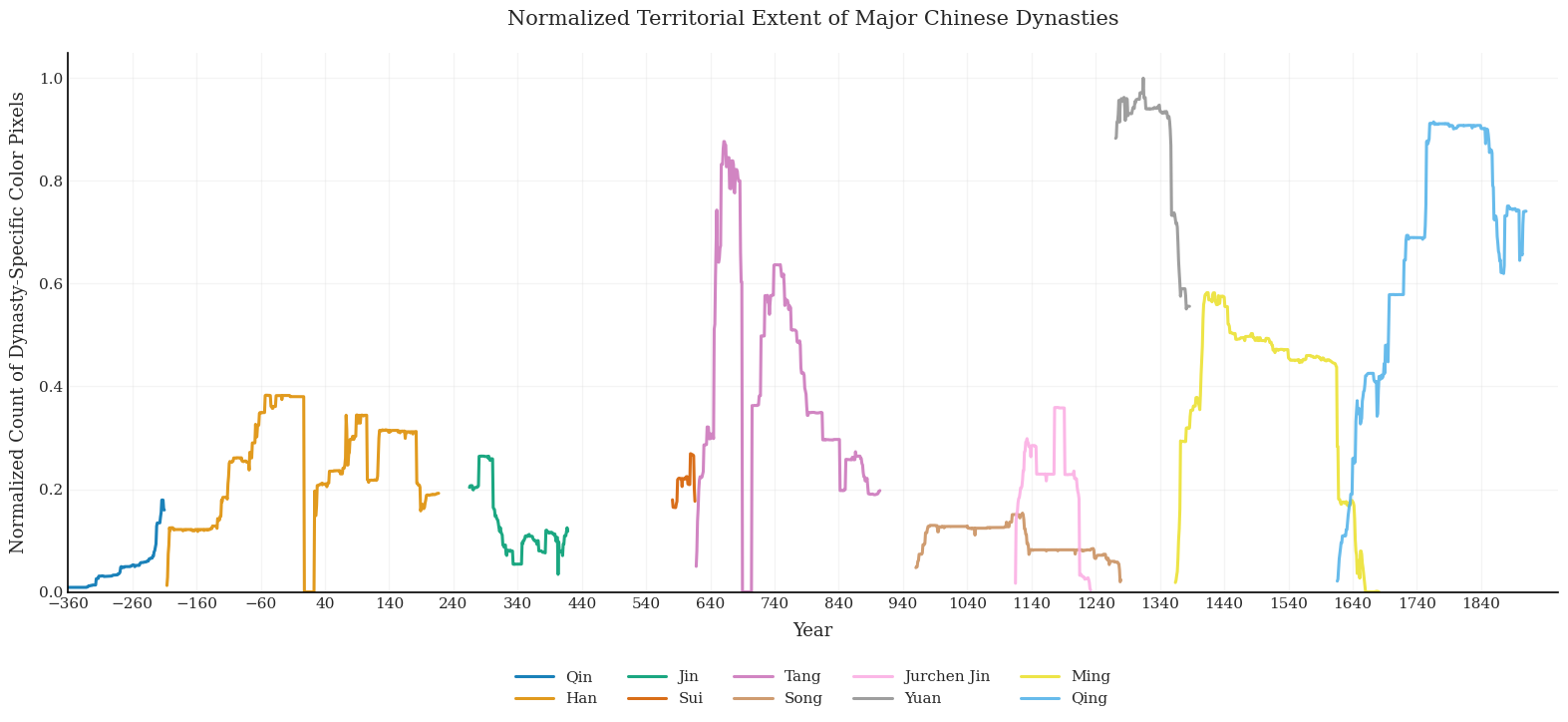}
    \caption{Normalized Territorial Extent of Major Chinese Dynasties (360 BCE–1912 CE). ATDE-extracted trajectories scaled via invariant water-body features (Appendix~\ref{Cross-video Scaling}) and normalized by the global maximum territorial extent (Yuan Dynasty = 1.0). The visualization enables comparison of both absolute territorial sizes and temporal patterns of expansion, stability, and decline across over two millennia of Chinese imperial history.}
    \label{fig:ATDE_full}
\end{figure}

Figure~\ref{fig:ATDE_full} presents all ten dynasties on a single plot after applying both scale normalization and relative normalization. Each trajectory now reflects the dynasty's territorial dynamics as a proportion of the global maximum, while maintaining spatial commensurability through water-body scaling. This visualization enables direct comparison of territorial stability, expansion rates, and collapse patterns across different historical periods and dynastic cycles.

Each trajectory now reflects the dynasty's territorial dynamics as a proportion of the global maximum

\section{Discussion and Limitations}

ATDE successfully extracts temporal trajectories from animated historical map videos, converting rasterized visualizations into structured time-series data. However, several important limitations should be considered when applying this tool.

\paragraph{Dependence on Source Quality.} The accuracy of extracted data is fundamentally constrained by the quality and historical accuracy of the source videos. Animated maps created for popular audiences often simplify complex territorial realities: Neglecting contested zones, tributary relationships, and gradual transitions in control. ATDE extracts what the video depicts, not necessarily what historical records support. Users should validate extracted trajectories against authoritative sources when using this data for research purposes.
\paragraph{Manual Calibration Requirements.} While the core pipeline is automated, ATDE requires manual specification of seed colors, HSV ranges, and filtering thresholds for each video. The RGB channel restrictions and Direct Neighbor Filtering parameters often need video-specific tuning to handle artifacts like ocean pixels or background noise. This requires user judgment, particularly when processing videos with inconsistent color schemes or visual styles.

\paragraph{Cross-Video Comparison Constraints.} Meaningful comparison across videos requires either a shared basemap (as in our Chinese dynasties case) or additional calibration steps. When videos use different projections, scales, or geographic extents, raw pixel counts cannot be directly compared. Our invariant-feature scaling approach (Appendix~\ref{Cross-video Scaling}) addresses zoom differences within a shared basemap but does not generalize to videos with fundamentally different cartographic foundations. Comparative trend analysis—examining patterns of expansion and contraction rather than absolute magnitudes—remains feasible across disparate sources, though with appropriate caveats about differing geographic contexts.

\paragraph{Future Directions.} Several extensions could enhance ATDE's capabilities: (1) automated parameter selection using machine learning, (2) semantic segmentation models to replace rule-based color detection, and (3) generalization to handle videos with multiple overlapping territories.

Despite these limitations, ATDE serves its intended purpose: providing a practical tool for extracting structured data from animated visualizations. It is best suited for educational demonstrations, preliminary data exploration, and analyzing relative trends. For example, even without absolute spatial calibration, ATDE enables comparison of how quickly different dynasties expanded, how long they maintained territorial stability, and whether collapse occurred gradually or abruptly—questions about temporal dynamics rather than absolute geographic extent.

\appendix
\section{Raw Frame Schematic}

\begin{figure}[H]
    \centering
    \includegraphics[width=.8\textwidth]{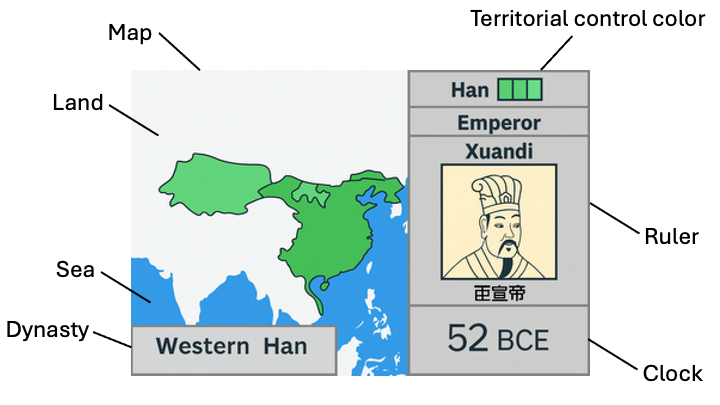}
    \caption{Schematic of a raw video frame illustrating the regions used in our analysis. 
    Only three components are retained for processing: the map (territorial colors), 
    the clock (used for temporal alignment), and the color legend encoding control status. 
    Other elements such as dynasty labels, portraits, and decorative graphics are discarded.}
    \label{fig:schematic}
\end{figure}

\section{Scale Normalization} \label{Cross-video Scaling}

Similarly, we can use the notion of invariant terrain features, specifically bodies of water (due to the reliable differentiation in RGB value composition), to account and correct for changes in the zoom, or "scale" of the maps. Video creators sometimes implement these inconvenient zoom changes in videos where less territory is "in play" and therefore less of the map is needed to understand the historical developments taking place. In order to interpret the sum of pixel values controlled by a dynasty meaningfully across numerous dynasty videos, we can utilize our uniquely invariant feature in this case of historical Chinese dynasties, the Bohai (and Yellow) sea. Despite the differences in coverage and features that our collection of videos displays, this gulf of water remains consistently featured throughout all of them.
\begin{table}[h!]
\centering
\begin{tabular}{|l|c|}
\hline
\textbf{Dynasty} & \textbf{Scale Factor} \\ \hline
Qin        & 0.113\\ \hline
Sui        & 0.622\\ \hline
Song       & 0.589\\ \hline
Jin        & 0.623\\ \hline
Ming       & 0.628\\ \hline
Qing       & 0.684\\ \hline
Yuan       & 0.698\\ \hline
Han        & 0.684\\ \hline
Tang       & 1.000 \\ \hline
Later Jin  & 0.632\\ \hline
\end{tabular}
\caption{Scale Factors for Individual Dynasties}
\label{tab:scale_factors}
\end{table}

If we can identify the sizes, in water pixels, of each video's Bohai and Yellow sea region, we can use those as yardsticks to determine the necessary scale factors. Before any calculations, we first need to identify the aforementioned region in order to start counting any "water" pixels. Using Matplotlib's "G-Input" feature, our code calls upon the user to distinguish this region by presenting screenshots from each dynasty's map and prompting the user to click on two coordinates which make a box around the sea region. Our code identifies the number of blue water pixels in each video's depiction of the seas and then divides each video's respective count of those pixels over our reference scale video (Tang Dynasty). These calculations equipped us with scale factors for every dynasty, which were applied when plotting to ensure accuracy in the mapping of territory won and lost. Table~\ref{tab:scale_factors} outlines the scale factors for each dynasty.

Dividing the raw number of pixels controlled by a dynasty by this scale factor, we arrive at a commensurable y-axis which allows us assemble Figure~\ref{fig:ATDE_full} and compare across dynasties.

\printbibliography

@article{Wilkinson2016FAIR,
  author       = {Wilkinson, Mark D. and Dumontier, Michel and Aalbersberg, IJsbrand Jan and Appleton, Gabrielle and Axton, Myles and Baak, Arie and Blomberg, Niklas and Boiten, Jan-Willem and da Silva Santos, Luiz Bonino and Bourne, Philip E. and Bouwman, Jildau and Brookes, Anthony J. and Clark, Tim and Crosas, Merc{\`e} and Dillo, Ingrid and Dumon, Olivier and Edmunds, Scott and Evelo, Chris T. and Finkers, Richard and Gonzalez-Beltran, Alejandra and Gray, Alasdair J.G. and Groth, Paul and Goble, Carole and Grethe, Jeffrey S. and Heringa, Jaap and 't Hoen, Peter A.C. and Hooft, Rob and Kuhn, Tobias and Kok, Ruben and Kok, Joost and Lusher, Scott J. and Martone, Maryann E. and Mons, Albert and Packer, Abel L. and Persson, Bengt and Rocca-Serra, Philippe and Roos, Marco and van Schaik, Rene and Sansone, Susanna-Assunta and Schultes, Erik and Sengstag, Thierry and Slater, Ted and Strawn, George and Swertz, Morris A. and Thompson, Mark and van der Lei, Johan and van Mulligen, Erik and Velterop, Jan and Waagmeester, Andra and Wittenburg, Peter and Wolstencroft, Katherine and Zhao, Jun and Mons, Barend},
  title        = {The {FAIR} Guiding Principles for scientific data management and stewardship},
  journal      = {Scientific Data},
  volume       = {3},
  pages        = {160018},
  year         = {2016},
  doi          = {10.1038/sdata.2016.18},
  note         = {URL: https://doi.org/10.1038/sdata.2016.18}
}

\end{document}